\documentclass[amsmath,amssymb, aps, prx, twocolumn, floatfix]{revtex4-2}

\usepackage{graphicx}
\usepackage{hyperref}
\usepackage{color}

\begin{document}
 
\title{Community detection forecasts material failure in a sheared granular material}

\author{Farnaz Fazelpour}
\author{Vrinda D. Desai}
\author{Karen E. Daniels}
\affiliation{Department of Physics, North Carolina State University, Raleigh, NC, USA}
  
\date{\today}
 
\begin{abstract}
The stability of a granular material is a collective phenomenon controlled by individual particles through their interactions. Forecasting when granular materials will undergo an abrupt failure is an ongoing challenge due to the intricate interactions between particles. Here, we report experiments on photoelastic disks undergoing intermittent stick-slip dynamics in a quasi-2D annular shear apparatus, with the evolving network of contact forces made visible via polarized light. We characterize the system by interpreting the interparticle forces as a multilayer network, and apply GenLouvin community detection to identify strongly correlated groups of particles. We observe that the community structure becomes increasingly volatile as the material approaches failure, and that this volatility provides a forecast that precedes what is detectable by considering the forces alone. We additionally observe that both weak and strong forces contribute to the strength of this forecast. These findings provide a new approach to detect patterns of causality and forecast impending failures. 
\end{abstract}

\maketitle

\section{Introduction}

Granular and other amorphous materials can rapidly lose stability and fail via the collective effects of local, interparticle interactions.  Forecasting when and where this will happen is both a challenge to our understanding of rigidity \cite{falk_dynamics_1998, tordesillas_force_2010, le_bouil_emergence_2014}, and also of practical importance in mitigating the human risk from natural disasters such as landslides \cite{intrieri_forecasting_2019, cascini_forecasting_2022} and earthquakes \cite{jordan_operational_2011}. In granular materials, interparticle interactions take the form of branching networks of frictional contact forces \cite{liu_force_1995, majmudar_contact_2005, hurley_quantifying_2016} that support the material. These networks provide stability \cite{behringer_physics_2018, liu_jamming_2010, henkes_rigid_2016,liu_spongelike_2021} but ultimately lose stability and give way to abrupt failure \cite{daniels_force_2008, tordesillas_force_2010}. Direct observations of grain kinematics and force chain networks in laboratory-scale granular material have led to many advances in both the understanding of how deformation and failure depends on particle scale interactions 
\cite{wang_microscopic_2018, ando_grain-scale_2012, yuan_creep_2023} as well as identifying  regions of high risk for idealized packings \cite{manning_vibrational_2011, cubuk_structure-property_2017}. These collective interactions imprint themselves on the vibrational modes of the material \cite{ohern_jamming_2003,wyart_rigidity_2005,owens_acoustic_2013}, changes in acoustic properties have provided a plausible route to forecasting the loss of rigidity \cite{tanguy_vibrational_2010, brzinski_sounds_2018, rouet-leduc_machine_2017} and dynamic precursors have also been observed in other soft, amorphous systems \cite{aime_microscopic_2018, ju_real-time_2023}.  However, characterizing the evolution of heterogeneous force networks through both space and time remains challenging as a means to predict when and where failure will occur in a granular material.

Recently, the collective nature of granular mechanics has motivated the use of network science techniques \cite{papadopoulos_network_2018} to analyze the arrangement and interactions of the particles. The aim is to utilize microscale information (interparticle contacts and forces) to identify the mesoscale structures that give rise to bulk behavior. Some of these tools analyze only the topology of force structure, regardless of strength of connections, for instance via contact loop or cycles \cite{smart_evolving_2008, arevalo_topology_2010, tordesillas_force_2010, pugnaloni_structure_2016, giusti_topological_2016}, or persistent homology \cite{kramar_persistence_2013, lim_topology_2017, nabizadeh_network_2023}.  In other studies, the connections are weighted by the interparticle force, allowing for analysis via clustering \cite{deng_lifespan_2022,bassett_influence_2012}, community detection \cite{walker_taxonomy_2012,tordesillas_revisiting_2013,bassett_extraction_2015,papadopoulos_evolution_2016}, or centrality \cite{kollmer_betweenness_2019}. Community detection has been used in several studies of granular materials to obtain insight to mesoscale structure of force chains \cite{bassett_extraction_2015, bassett_influence_2012} including their evolving characteristics under compression \cite{papadopoulos_evolution_2016}. Network science techniques have been successful even at geophysical length scales, where individual grain positions cannot be resolved \cite{tordesillas_spatiotemporal_2021,singh_spatiotemporal_2020,desai_forecasting_2023}.
Furthermore, measures of centrality have seen mixed success \cite{berthier_forecasting_2019, pournajar_edge_2022} in identifying likely failure locations in disordered lattice-like structures.

\begin{figure}
\centering 
\includegraphics[width=1\linewidth]{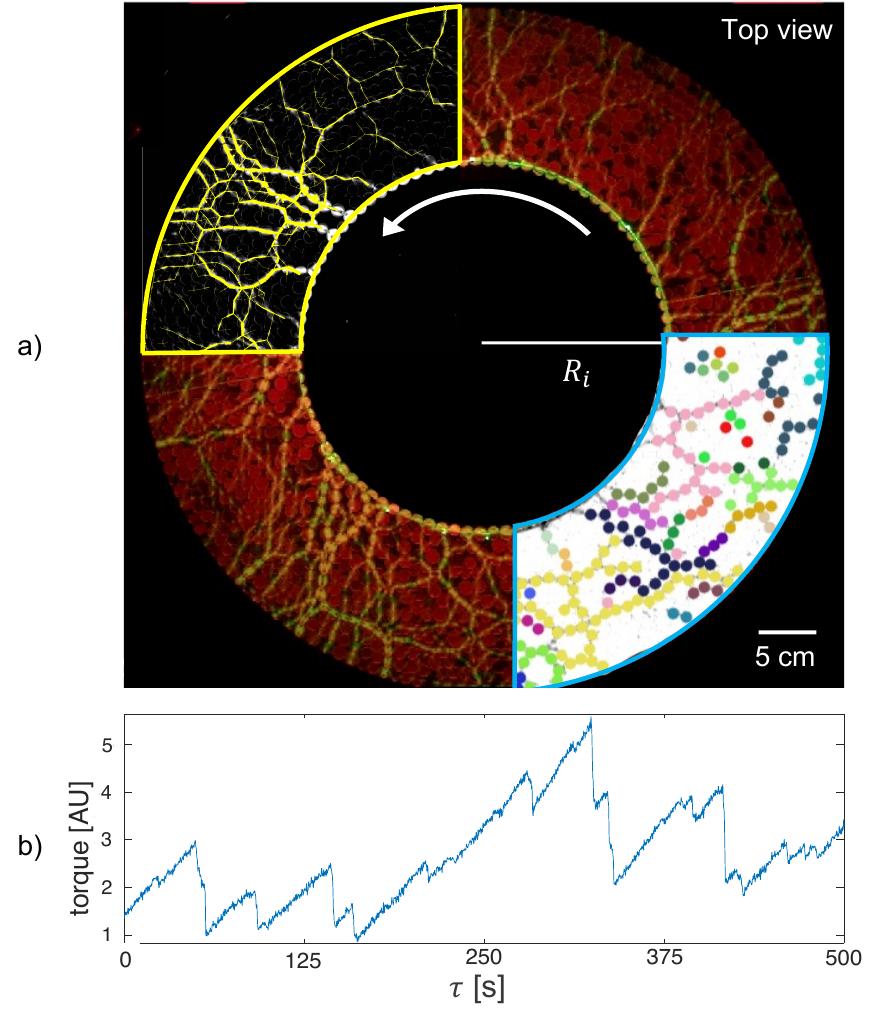} 
\caption{(a) Top view of annular Couette experiment filled with flat photoelastic particles. The particle positions are visualized in unpolarized red light, and the interparticle forces are measured using polarized green light (brighter corresponds to more force). The yellow segment (top left) shows an example weighted contact network;  the width of each edge is proportional to $G^2$ (for all contacts above a threshold). The blue segment (bottom right) shows an example community structure;  particles of the same color are in the same community and single-particle communities are not shown. 
(b) Sample torque data over $\tau=500$ seconds, showing typical variability in stick-slip magnitude and duration.}
\label{fig:whole_system} 
\end{figure}

Motivated by these many successful studies, this paper evaluates the utility of community detection techniques \cite{fortunato_community_2010, porter2009communities, kivela_multilayer_2014, boccaletti_structure_2014}  as a means to identify strongly-correlated clusters of particles within a sheared granular material contained within a quasi-two-dimensional granular experiment undergoing stick-slip failure (see Fig.~\ref{fig:whole_system}). A chief advantage of these methods is that they allow for the inclusion of temporal evolution of the networks, by connecting a sequence of strain steps into a {\it multilayer network}.  To achieve this, we write each layer as a set of nodes (tracked particles) connected by edges (weighted by interparticle force, measured using photoelasticity \cite{daniels_photoelastic_2017, abed_zadeh_enlightening_2019}). The nodes are connected to themselves, sequentially, across the series of layers, one layer for each image within a video of a stick-slip event. Using GenLouvin modularity maximization \cite{mucha_community_2010,lucas_g_s_jeub_generalized_2011}, we uncover the mesoscale patterns that lead to stick-slip granular failure. This method partitions the networks into communities of nodes that are more strongly connected to the other nodes within their community, than they are to those in other communities.  We observe that the communities  detected in this way become more volatile in the time leading up to failure, reminiscent of the clustered regions observed by \citet{le_bouil_emergence_2014} using diffusing wave spectroscopy. 
Furthermore, while the sensitivity for triggering on the precursor signal is strongest  when  both pre- and post-failure data is included in our analysis, the signal is also present using  only pre-failure data. This suggests that community detection is a promising route to forecasting the time of failure. Interestingly, weighting the contributions from either the largest or highest-force clusters more strongly does not improve the precursor signal, a finding that reiterates the importance of even weak force chains in stabilizing/destabilizing the collective network \cite{cates_jamming_1998, liu_spongelike_2021}.

\section{Methods} 

\subsection{Experiment}

Our apparatus is a quasi-2D annular shear cell (see Fig.~\ref{fig:whole_system}, details provided in \cite{brzinski_sounds_2018}). A single layer of disks rests on an acrylic substrate; these particles are confined between two walls: a rough inner wheel with radius $R_i=15$~cm and a smooth outer wall with radius $R_o=28$~cm. The inner wall is rotated by a motor ($0.035$ ${\mathrm{rad}}/{\mathrm{sec}}$) connected to the axle via a torsional spring (stiffness $0.85$ N$\cdot$m/rad) to generate intermittent stick-slip failure events. The motor loads the spring at a constant rate, at first compressing the spring rather than rotating the inner wall. When the torque surpasses the strength of the granular material, a slip occurs, the spring relaxes, the torque drops rapidly, and the inner wheel rotates forward.  The $\sim 1700$ disks are bidisperse, with equal populations of diameter $d_S = 0.9$~cm and $d_L = 1.1$~cm, both with thickness $6.3$~mm. 

We imaged the dynamics of the particles and forces via an Edgertronic high speed camera ($816 \times 1280$ resolution, $200$~Hz frame rate). The dataset we analyze in this paper comprises $23$ videos, each of duration $\sim 15$ seconds ($L=2995$ frames) and containing one stick period followed by one slip event, drawn from a continuous run during which such events arose intermittently. These 23 events were extracted from this run to center each failure event within a single video. We tracked the location of each of $N \sim 500$ particles in each image using the Blair-Dufresne particle-tracking algorithm \cite{BD} on positions determined via Hough transformation \cite{HT}, and determined the neighbors based on the interparticle separation distance being less than $1.1$ times the sum of the two particle radii. This tolerance generously accounts for small imperfections in locating particle centers. Particles do not enter or leave the imaging region during the stick period; during the slip period particle tracking is terminated once the particles leave the imaging region.

We quantify the interparticle forces via the photoelastic response of the Vishay PhotoStress material PSM-4 (bulk modulus $E=4$~MPa, density $\rho=1.06$~g/cm$^3$). All particles are located between two circular polarizers of opposite handedness, and the observed pattern of optical fringes depends quantitatively upon the imposed stress on the particles \cite{daniels_photoelastic_2017, abed_zadeh_enlightening_2019}. Due to limitations in the image resolution required to obtain high-speed videos, we quantify the forces using the $G^2$ technique, which measures the particle-scale gradient in image intensity; this has been found to be approximately proportional to the force magnitude \cite{howell_stress_1999, zhao_particle_2019}. To formulate the multilayer network for each stick-slip event, we locate and track all particles in the frame, identify their neighbors, and assign each of these contacts a weighted magnitude corresponding to the measured $G^2$ averaged for the two particles. We include data for all contacts above a threshold value selected to balance over- and under-counting. A sample network is shown in yellow in Fig.~\ref{fig:whole_system}a, with the thickness of the edges proportional to $G^2$.  

\subsection{Community Detection}

Multilayer networks comprise a series of single networks (layers) that are then connected to each other across the layers as well as within them. As such, they have been used in engineering, social, and biological contexts to interpret interactions both between and within those layers \cite{kivela_multilayer_2014, boccaletti_structure_2014, papadopoulos_evolution_2016}.  While individual layers can represent data of different types, layers which describe only a single type of data -- as we will use here -- provide a way to directly quantify spatiotemporal dynamics. We therefore construct a multilayer network from a time-series of single-layer networks in which each node corresponds to a particle whose interparticle interactions change during the lead-up and aftermath of a failure event. Each multilayer network is centered on the time of failure as the middle layer.

As done previously for the analysis of granular dynamics \cite{papadopoulos_evolution_2016}, the edges (connections) between nodes (particles) are described by an adjacency matrix $\mathcal{A}$, populated with elements
\begin{equation}
    \mathcal{A}_{ijt} = 
\begin{cases}
    f_{ijt}& \text{if nodes $i$ and $j$ are in contact} \\
    0,              & \text{otherwise}
\end{cases}
\end{equation}
where $f_{ijt}$ is the interparticle force between two particles $i$ and $j$ at layer (time) $t$. Here, force is measured using the $G^2$ photoelastic method described above, and due to force balance being approximately satisfied in quasi-static flows, we assure that $\mathcal{A}$ is symmetric. 
Particle identities ($i,j$) are fixed for the duration of an individual event, so that the  multilayer network contains both the topological structure of force chains as well as  their temporal dynamics, allowing us to characterize which features give rise to stick-slip failure.

Motivated by the observation that the rigidity of a granular system arises at the mesoscale, within clusters of particles \cite{henkes_rigid_2016, liu_spongelike_2021}, we analyze this multilayer network using a community-detection technique known as modularity maximization \cite{newman_finding_2004,newman_modularity_2006}. We use the GenLouvain algorithm by Mucha et al. \cite{mucha_community_2010, lucas_g_s_jeub_generalized_2011} to partition the nodes into multilayer communities  $c$ for each of the 23 events, via maximizing the modularity $Q$: 
\begin{equation}
Q = \frac{1}{\mu}\sum_{ijtm} [ (\mathcal{A}_{ijt}-\gamma \mathcal{P}_{ijt})\delta_{tm}+\omega\delta_{ij}] \delta (c_{it},c_{jm})
\label{eq:multilayer_modularity}
\end{equation}
Here $c_{it}$ identifies the community to which a particle (node) $i$ belongs at layer $t$ and the algorithm heuristically adjusts all community assignments to maximize the value of $Q$. The quality factor (given within the square brackets) for particles $i,j$ at two adjacent layers $t,m$ (where $m=t+1$) compares the adjacency matrix 
$\mathcal{A}$ to a null model for all particles assigned to the same community (via the Kronnecker  $\delta (c_{it},c_{jm})$).

The quality factor compares $\mathcal{A}$ to a null model defined as follows.  In spatially-embedded networks such as ours, particles can only interact with their nearest neighbors. As such, the commonly-used Newman-Girvan null model \cite{newman_modularity_2006,newman_finding_2004} would be non-physical in this case.  Our analysis instead uses the geographical null model $\mathcal{P}$ \cite{bassett_robust_2013, bassett_extraction_2015, papadopoulos_evolution_2016}, with a small modification to use the observed physical distances between particles as the definition of being neighbors; this allows us to be less sensitive to any missing contacts not found during the photoelastic analysis.
Our null model is given by $\mathcal{P}_{ijt}=\bar{f}_{t} B_{ijt}$ where  $B$ is a binary adjacency matrix which is nonzero only when the particles are neighbors (as defined above, based on particle separation). The weight $\bar{f}_{t}$ is the expected weight which is the average interparticle force in that layer.
It is important to allow for a time-dependent null model since the average force within the material varies strongly over time, as illustrated by the torque measurements in Fig.~\ref{fig:whole_system}. In practice, we accomplish this by normalizing each layer $\mathcal{A}_t$ by the average interparticle force for that layer and then set $\bar{f}_{t}=1$. 
The null model for the multilayer aspect is that each particle is connected to itself throughout layers, given by the Kronnecker $\delta_{ij}$. The number of communities into which the system is partitioned is controlled by the spatial resolution parameter $\gamma$ and the interlayer coupling $\omega$; in both cases, the larger the value, the larger the number of communities (and the smaller those communities). We follow the methods of \citet{papadopoulos_evolution_2016} in order to select $\gamma=1$ and $\omega=0.5$ (see Results, below) for all analyses shown here. By convention, the quality factor is normalized by  $\mu=\sum_{jt}(\sum_{i}\mathcal{A}_{ijt}+\sum_{m}\omega)$.

\section{Results} 

To reduce the noise in interparticle force measurements, we generated an adjacency matrix $\mathcal{A}$ for each of the 23 runs, and applied a boxcar average over 5 frames (thereby reducing the temporal resolution).
This results in a multilayer force network with $599$ layers separated by $0.025 sec$ in time. Using the GenLouvain multilayer community detection process described above, we detect communities by maximizing modularity (Eq.~\ref{eq:multilayer_modularity}); a sample outcome is shown in Fig.~\ref{fig:whole_system}a for a single layer. This determines a set of communities, each shown in a unique color, for which the particles inside that community are more strongly connected to each other than they are to neighbors not in their community. Through the GenLouvin optimization process, particles can switch communities during each subsequent layer, but are biased to remain connected through the interlayer coupling $\omega$. 

\begin{figure}
\centering 
\includegraphics[width=0.9\linewidth]{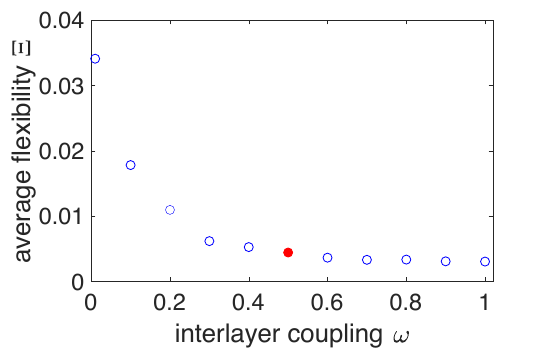} 
\caption{Average flexibility $\Xi$, measured using Eq.~\ref{eq:flex_tot}, for various values of the interlayer coupling $\omega$.  The chosen value $\omega=0.5$ is marked in red.
}
\label{fig:flexibility} 
\end{figure}

The resolution parameter $\gamma$ controls the size of the communities, which we take to be a constant rather than allowing it to vary between layers. Following on prior work in the detection of communities in granular systems, \cite{bassett_extraction_2015,papadopoulos_evolution_2016}, we select $\gamma=1$. The corresponds to comparing the interparticle forces to the average force.
The other parameter that needs to be selected is the interlayer coupling $\omega$, which we also take to be a constant. This parameter controls how strongly nodes are connected to themselves in adjacent layers. To identify the optimal value that allows for enough community flexibility, we follow a similar analysis as was done in \citet{papadopoulos_evolution_2016}. To quantify how often nodes switch communities between layers, we measure the flexibility $\xi_i$ of each node:
\begin{equation}
\xi_{i}=\frac{g_{i}}{L-1}
\label{eq:flex}
\end{equation}
where $g_{i}$ is the number of times that particle $i$ switches communities throughout that multilayer network, and $L$ is the total number of layers. Accordingly, the average flexibility for  all particles is given by
\begin{equation}
\Xi=\frac{1}{N}\sum_{i}\xi_{i}
\label{eq:flex_tot}
\end{equation}
where $N$ is the total number of nodes (particles), and $\Xi$ represents the flexibility of the whole network. We implement community detection for a range of interlayer coupling values from $\omega=0.01$ to $\omega=1$, and measure the average particle flexibility for each value.  Fig.~\ref{fig:flexibility} shows this average flexibility for an example event. For the present study, we select $\omega=0.5$ since this is the smallest value for which flexibility has largely plateaued.

Our main dataset contains a list of identified communities for the chosen $(\gamma, \omega)$. These detected communities vary in size and strength throughout the stick-slip process; in all cases we neglect singleton communities (containing only one particle) in our calculations. Using these communities, we seek to identify consistent patterns in community structure and dynamics that allows us to both understand the process of failure, and possibly forecast when it will occur. Thus, we perform our analysis using the full multilayer network (both pre- and post-failure data) for each event, and also using only the  pre-failure data.

We focus on three measures which characterize community structure at each time step. We measure {\it community size} $s_t^c$ as the number of particles in community $c$ at layer $t$, and {\it community strength} $\sigma_t^c$ as 
\begin{equation}
\sigma_{t}^c=\frac{\sum_{ij\in c}\mathcal{A}_{ijt}}{s_t^c} 
\label{eq:strength}
\end{equation}
the average edge weight (interparticle force magnitude) in community $c$ at layer $t$. Motivated by the ubiquity of critical-point fluctuations near phase transitions --- even for nonequilibrium systems --- we additionally characterize {\it community volatility} as a measure of how  much the community assignments change from layer to layer. For each community $c$ the volatility
\begin{equation}
\nu^{c}_{t}=\frac{\lvert c_t \bigtriangleup c_{t+1}\rvert}{\lvert c_t \cup c_{t+1}\rvert}
\label{eq:vol_c}
\end{equation}
is measured at layer $t$. The numerator $\lvert c_t \bigtriangleup c_{t+1}\rvert$ is the number of particles in community $c$ at $t+1$ but not at layer $t$ (particles that leave community $c$). The denominator $\lvert c_t \cup c_{t+1}\rvert$ is the total number of particles in community $c$ at either $t$ or $t+1$. The average volatility of all communities in each layer is given by 
\begin{equation}
 \bar{\nu}_{t}= \frac{1}{n_t}\sum_{c} \nu^{c}_{t}
 \label{eq:nu}
\end{equation}
where $n_t$ is the number of communities at layer $t$. This quantity represents how much the community structure is changing during each timestep, with larger values indicating more particles are switching communities. 

\begin{figure}
\centering 
\includegraphics[width=1\linewidth]{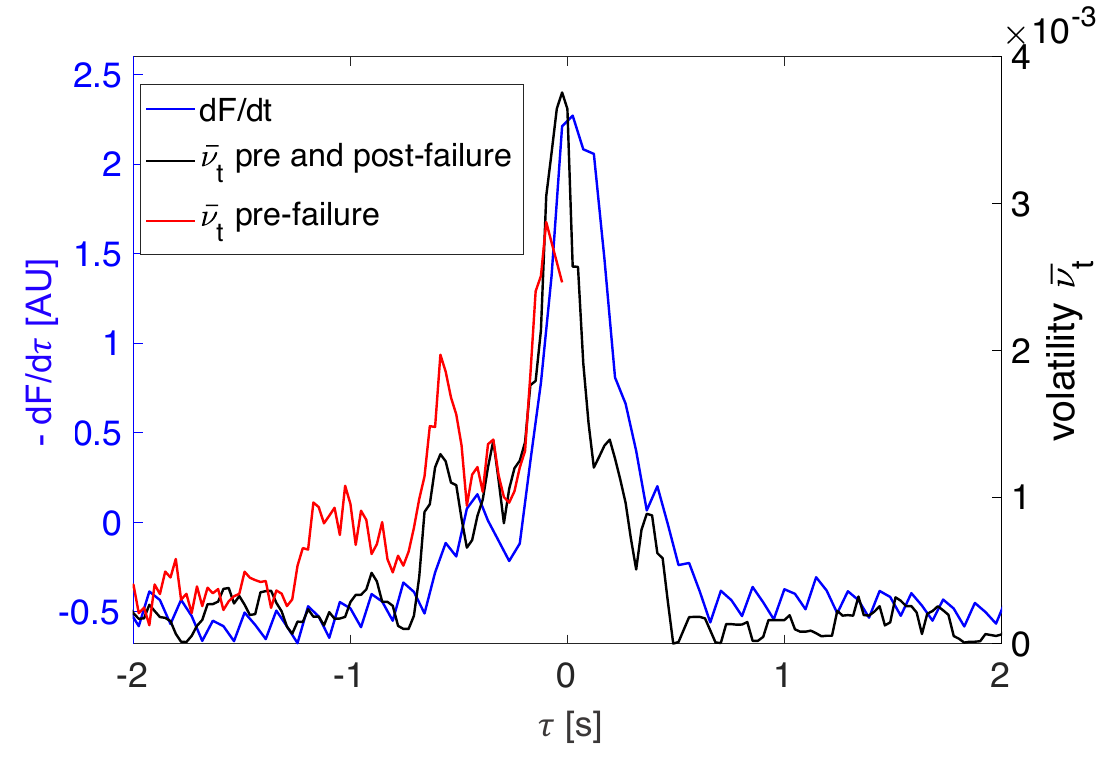} 
\caption{Comparison of force and volatility dynamics for a representative event. Blue line: the rate of change of the total force measured by $G^2$, in arbitrary units; the slip event initiates at the maximum of this curve. Black line: mean volatility $\bar{\nu}_t$ (Eq.~\ref{eq:nu}), calculated using both pre- and post-event data. Red line: mean volatility $\bar{\nu}_t$ (Eq.~\ref{eq:nu}), calculated using only pre-event data. All data is shown with additional smoothing over a moving window of $5$ data points to assist in visualizing the trends. 
}
\label{fig:volatility} 
\end{figure}

The total force within the system  (measured by summing $G^2$ values)  rises during the stick phase of the dynamics ($\tau < 0$~s), and falls during the slip phase  ($\tau > 0$~s) as contacts are mobilized and then release the built-up forces. Even though the system is globally rigid, there are nonetheless small changes in the interparticle forces during the stick phase, and these changes cause variability in the assignment to communities. Fig.~\ref{fig:volatility} shows a graph of the numerical values calculated from these dynamics.
We observe that the total force $F$ within the layer is fluctuating prior to failure and that the value only rises above statistical fluctuations for $\tau \gtrsim  -0.6$~s. Meanwhile,  $\bar{\nu}_t$, whether measured over only the pre-failure data or for the full event, exhibits a rise above the background for $\tau \gtrsim -1$~s. The peak in the rate at which interparticle forces are increasing arrives when force chains start breaking; the highest value is therefore when the main slip event occurs. Variations in this rate suggest that force chains are fluctuating prior to failure, and that network is changing at the community scale due to force fluctuations prior to major force chains breaking.

\begin{figure}
\centering 
\includegraphics[width=1\linewidth]{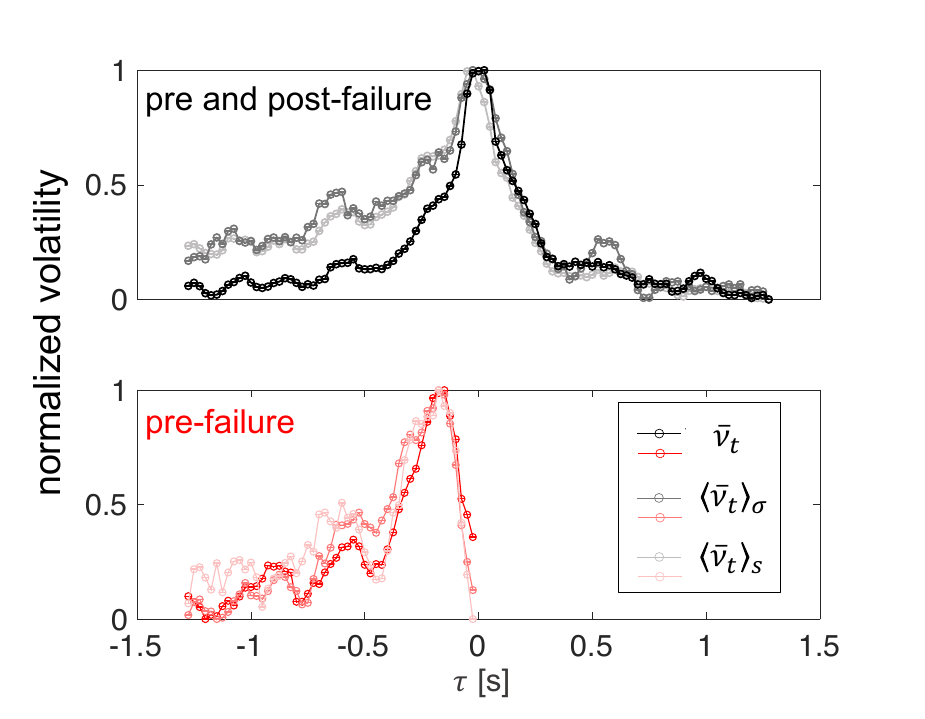} 
\caption{Comparison of volatility measurements using both pre- and post-failure data vs. pre-failure data only, averaged over all communities and over all 23 events. The averages are taken in three different ways, to understand the controls on this process. The darkest line is the simple average $\langle \bar{\nu}_t \rangle$ taken over all 23 events (recentered to $\tau=0$), normalized to a minimum value of 0 and a maximum value of 1. The medium darkness line takes the same average, but each community's contribution is weighed by its strength $\sigma_c$ (see Eq.~\ref{eq:sigweight}). The lightest line has each community's contribution weighted by its size $s_c$  (see Eq.~\ref{eq:sizweight}). }
\label{fig:average} 
\end{figure}

To generalize from this example, we consider the average volatility $\langle \bar{\nu}_t\rangle$, for which Eq.~\ref{eq:nu} is averaged over all $23$ slip events, where each time-axis ($\tau$) is recentered to have its failure time at $\tau=0$~s and $\langle \cdot \rangle$ denotes the average over all events. Since it is possible that stronger communities contribute more volatility, we also consider averages weighted by community strength
\begin{equation}
 \langle \bar{\nu}_{t} \rangle_\sigma =\left\langle \frac{1}{n_t} \sum_{c} \sigma^{c}_{t} \nu^{c}_{t} \right\rangle
 \label{eq:sigweight}
\end{equation}
Similarly, we calculate the average volatility, weighted by the size of each community
\begin{equation}
 \langle \bar{\nu}_{t} \rangle_s =\left\langle \frac{1}{n_t} \sum_{c} s^{c}_{t} \nu^{c}_{t} \right\rangle
 \label{eq:sizweight}
\end{equation}

Fig.~\ref{fig:average} compares the results of these three averages, taken over either the pre- and post-failure data (black lines), or over the pre-failure data only (red lines). We observe that in all six cases, the average volatility rises during the lead-up to failure. This rise occurs even when we do not include any post-failure data (panel b), and the rise becomes apparent for $\tau > -1$~s, much earlier than is apparent from force data ($dF/d\tau$) alone. As such, it shows promise for forecasting the time of failure events. it is interesting that excluding post-failure data caused the peak to move earlier in $\tau$ because the number of communities increased, including singleton communities and communities that exist only in one layer, which are not considered in volatility measurements.

By comparing the results of Fig.~\ref{fig:average}a, depending on whether or not size and strength are taken into account, we can examine whether the volatility is primarily present in either strong or large communities. For example, it is possible that the strongest force chains are responsible for slip events, for instance via buckling \cite{tordesillas_force_2010}.
Instead, we observe that the strength-weighted volatility $\langle \bar{\nu}_{t} \rangle_\sigma$ and size-weighted volatility $\langle \bar{\nu}_{t} \rangle_s$ in fact under-perform the straight average (shown as the darkest line).   This  implies that small and weak communities play a significant role in stabilizing the system, and volatility in those regions is contributing to improving the forecast of the impending time of failure. This observation is consistent with the result of \citet{henkes_rigid_2016, liu_spongelike_2021}, where both rigid and floppy regions -- detected through a mesoscale stability analysis -- are observed to both contain strong force chains, and to be of highly-variable sizes. 

Ensemble averages, however, do not speak to whether or not a forecast of failure would have been possible during each specific event. To quantitatively evaluate our success at making forecasts, we turn to considering the changes in volatility $\nu_{t}$ for individual events, monitored along their route to failure. Again we consider our data using either pre- and post-failure data, or pre-failure data alone. 

Our procedure for making a forecast considers each layer prior to failure, recording the change in volatility with respect to the average volatility for all prior times within the event interval.  If $\bar{\nu}_t > M\sum^{t-1}_{i=1}\bar{\nu}_i$ we forecast a possible future failure, where $M$ sets the size of the threshold. For various values of $M$, we observe how many of the 23 events would have recorded a  true positive (TP), false positive (FP), true negative (TN), or false negative (FN) under that criterion. Due to the noisiness of individual events' volatility measurement (see Fig.~\ref{fig:volatility}), we examine a window from $1 s$ before failure to the last step before failure in making these assignments.  $\bar{\nu}_t > M\sum^{t-1}_{i=1}\bar{\nu}_i$ in this window is (TP), outside this window is (FP) and $\bar{\nu}_t < M\sum^{t-1}_{i=1}\bar{\nu}_i$ in this window is (FN), outside this window is (TN).

\begin{figure}
\centering 
\includegraphics[width=1\linewidth]{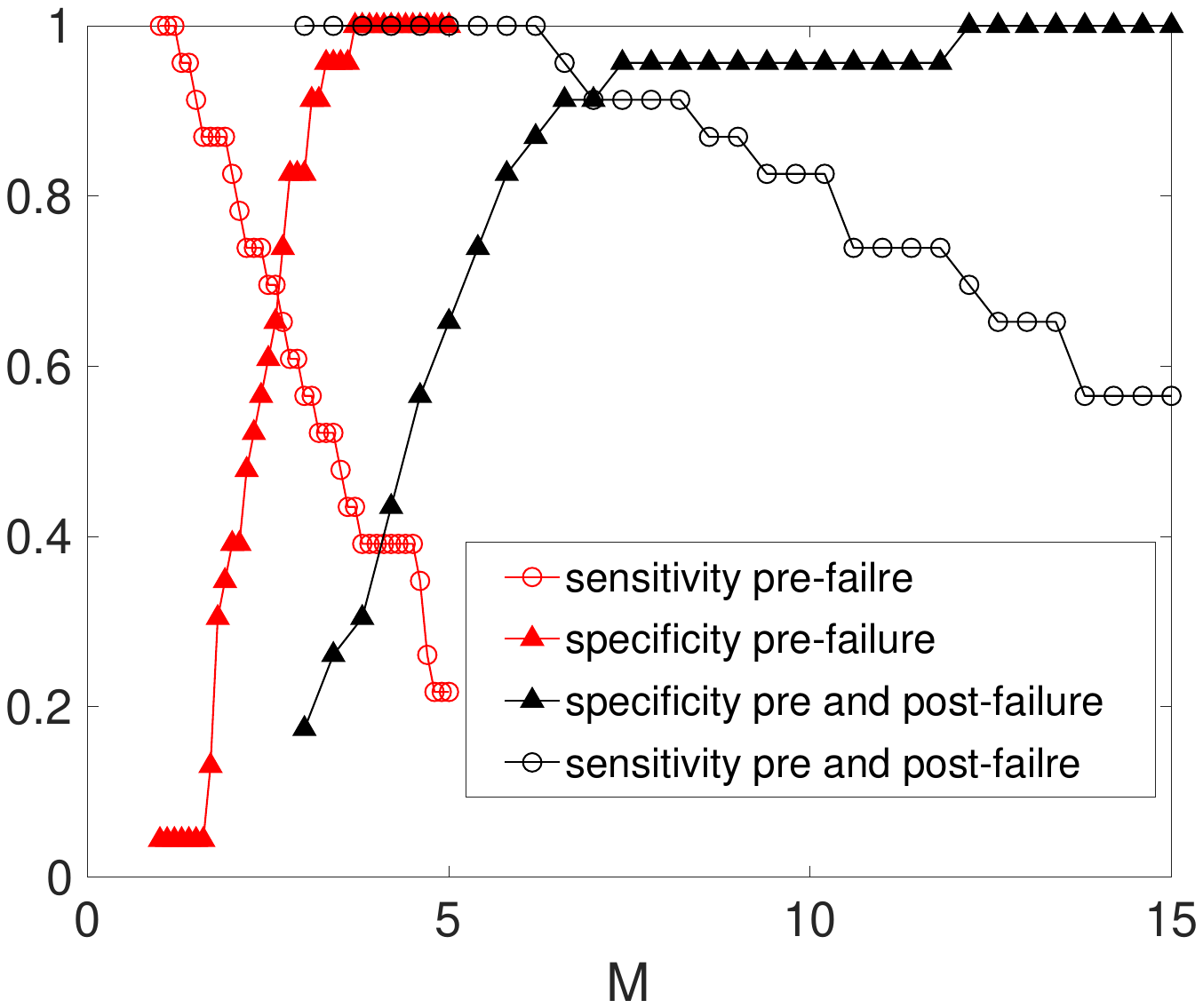} 
\caption{Measured sensitivity (open circles) and specificity (solid triangles) for the choice of threshold $M$. Black symbols/lines: pre- and post-failure data. Red symbols/lines: pre-failure data only. The crossing points are at $91\%$ ($M=7.0$) and $65\%$ ($M=2.7$), respectively.
}
\label{fig:sensitivity_specificity} 
\end{figure}

The success of a forecasting algorithm can be described by its {\it sensitivity} $=\mathrm{TP}/(\mathrm{TP}+\mathrm{FN})$ (probability of correctly forecasting an observed failure event)  and {\it specificity} $=\mathrm{TN}/(\mathrm{TN}+\mathrm{FP})$ (probability of forecasting no failure when material doesn't fail). Selecting a particular value of the threshold $M$ involves making a trade-off between these two considerations.  Fig.~\ref{fig:sensitivity_specificity} quantifies these trade-offs, as well as how they differ depending on whether or not we include post-failure data within the analysis.  As expected, increasing the threshold $M$ decreases the  sensitivity, but increases the specificity. The particular choice of $M$ reflects the competition between forecasting all possible failures, without introducing too many false positives. We find that correct forecasts are possible for the majority ($65\%$) of events, even without including post-failure data. With the inclusion of that data, the success rate rises to $91\%$, and the use of a higher threshold is possible.

\section{Discussion \& Conclusions} 

This study takes place in the context that the evolution of force chains provides a complex system for which the proximity to Coulomb failure is difficult to quantitatively establish at the particle-scale, and for which the mesoscale is known to provide important controls on the stability. Intriguingly, network science tools have a long track record of providing insights for noisy data for which the true interactions are only estimated, in part because network topology is such a strong control. 

In this paper, we have presented a new description of the evolution of the configurations of interparticle forces, drawing on the tools of network science, which highlights the importance of the volatility in the detection of communities of strong force chains. 
We observe that variations in the community structure significantly increase (via small force fluctuations) during the interval of time leading up to failure. This volatility defines a precursor state that can be used to forecast the failure of the material at the bulk scale, and is successful when used on  pre-failure data alone. The inclusion of post-failure data improves the success of the method, and also allows us to investigate the role of community size and strength. We find that weighting the volatility by size or strength does not improve the forecasting, and this finding demonstrates that weak and small communities play a crucial role in material stability, and hence in forecasting failures.

It is notable that similar tools are being applied to predict  landslides using remote sensing data \cite{tordesillas_spatiotemporal_2021, singh_spatiotemporal_2020}. We have recently observed \cite{desai_forecasting_2023} that community detection of the type described here can be applied at the landscape scale. Intriguingly, in that case the community volatility decreases rather grows on approach to failure, with the caveat that no forces have been measured, only strain fields.

\begin{acknowledgments}
This work was supported by NSF grants DMR-2104986, DMR-1206808, ICER-1854977, and the James S. McDonnell Foundation. We are grateful to Mason Porter, Katie Newhall, and the PREEVENTS collaboration for helpful discussions.
\end{acknowledgments}

\bibliographystyle{apsrev}
\bibliography{r}

\end{document}